\documentclass[conference, letterpaper]{IEEEtran}

\usepackage{mathtools, amssymb}
    \def\A{\mathsf A}
    \def\C{\mathsf C}
    \def\G{\mathsf G}
    \def\T{\mathsf T}
    \def\EE{\mathbb E}
    \newtheorem{theorem}{Theorem}
    \newtheorem{lemma}[theorem]{Lemma}
    \newtheorem{question}[theorem]{Question}
    \newtheorem{corollary}[theorem]{Corollary}
    \newtheorem{definition}[theorem]{Definition}
    \newtheorem{proposition}[theorem]{Proposition}

\usepackage{caption, subcaption, tikz-cd}
    \usetikzlibrary{decorations.pathmorphing}

\usepackage{snaptodo}
    \snaptodoset{margin block/.style={font=\tiny}}

\usepackage[colorlinks, allcolors=purple!50!blue!90!black]{hyperref}

\usepackage{autonum}

\begin{document}

                                 \title
                       {On Counting Subsequences
                 \\ and Higher-Order Fibonacci Numbers}
                                    
                                \author
                    {\IEEEauthorblockN{Hsin-Po Wang}
                           \IEEEauthorblockA
      {Department of Electrical Engineering and Computer Sciences
            \\ Simons Institute for the Theory of Computing
             \\ University of California, Berkeley, CA, USA
                     \\ Email: simple@berkeley.edu}
                                  \and
                    \IEEEauthorblockN{Chi-Wei Chin}
                           \IEEEauthorblockA
                  {Apricob Biomedicals Co Ltd, Taiwan
                    \\ Email: qin@quantumsafe.dev}}
                                    
                               \maketitle

\begin{abstract}\boldmath
    In array-based DNA synthesis, multiple strands of DNA are
    synthesized in parallel to reduce the time cost from the sum of
    their lengths to the length their shortest common supersequences.
    To maximize the amount of information that can be synthesized into
    DNA within a finite amount of time, we study the number of unordered
    sets of $n$ strands of DNA that have a common supersequence whose
    length is at most $t$.  Our analysis stems from the following
    connection: The number of subsequences of A C G T A C G T A C G T
    ...  is the partial sum (prefix sum) of the fourth-order Fibonacci
    numbers.
\end{abstract}

\section{Introduction}

\begin{figure}
        \centering
        \pgfmathsetseed{20240120}
        \begin{tikzpicture}
            \def\0{\A} \def\1{\C} \def\2{\G} \def\3{\T}
            \def\letter#1{
                \pgfmathtruncatemacro\modfour{mod(#1, 4)}
                \csname\modfour\endcsname
            }
            \def\strand[#1]#2{
                \tikzset{xshift=2.28cm}
                \draw (0, -1/3) node [inner sep=0] {master};
                \draw [#1] (1.25, -1/3) node {#2};
                \foreach \t in {0, ..., 11} {
                    \draw (0, \t/3) node (m\t) {$\letter{\t}$};
                }
                \def\y{0}
                \foreach \t in {0, ..., 11} {
                    \pgfmathsetmacro\rnd{rnd}
                    \ifdim \rnd pt > 0.5pt
                        \xdef\y{\the\numexpr\y+1\relax}
                        \draw [#1]
                            (1, -0.3+\y/2) node (x\t) {$\letter{\t}$}
                            (x\t) ++(.1, 0) -- ++(.1, 0)
                            +(0, .2) -- +(0, -.3)
                        ;
                        \draw [->] (m\t) -- (x\t);
                    \fi
                }
            }
            \strand[red]{$x^1$}
            \strand[blue]{$x^2$}
            \strand[teal]{$x^3$}
            \strand[brown]{$x^4$}
        \end{tikzpicture}
        \caption{
            Array-based DNA synthesis: At time $s = 1$, we ask $x^1,
            x^2, x^3, x^4$ whether they want $M_1 = \A$; only $x^2$ says
            yes.  At time $s = 2$, we ask whether they want $M_2 = \C$,
            and $x^3, x^4$ say yes.  The same process repeats until we
            ask whether they want $M_{12} = \T$, and $x^1$ says yes.
        }                                             \label{fig:master}
    \end{figure}
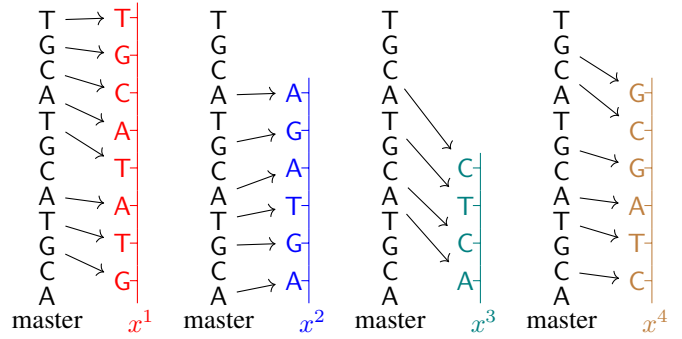

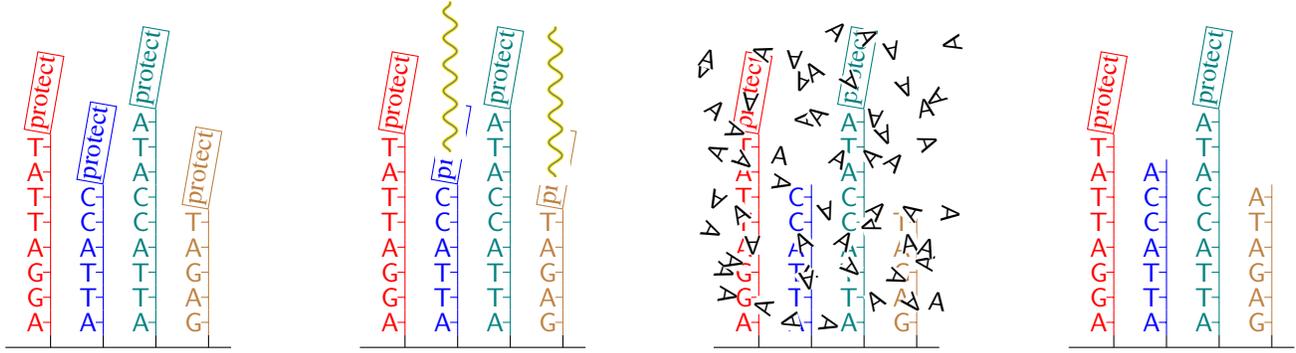
\begin{figure*}
        \centering
        \def\strand[#1]#2#3{
            \tikzset{xshift=0.7cm}
            \draw (0.2, 1/6) -- +(0, -1/6);
            \xdef\y{0}
            \begin{scope}[#1]
                \strandaux #2;
                #3
            \end{scope}
        }
        \def\strandaux#1#2;{
            \if#1;
            \else
                \xdef\y{\numexpr\y+1\relax}
                \draw
                    (0, \y/3) node (X) {$\csname#1\endcsname$}
                    (X) ++(0.1, 0) -- ++(0.1, 0)
                    +(0, 1/6) -- +(0, -1/6)
                ;
                \strandaux#2;;
            \fi
        }
        \def\prot{
            \draw
                (0.2, \y/3+1/6) node
                [above right, rotate=80, draw, inner sep=1]
                {protect}
            ;
        }
        \def\light{
            \draw
                (0.2, \y/3+1/6) node
                [above right, rotate=80, draw, inner sep=1]
                {protect}
            ;
            \fill [white]
                (-0.1, \y/3+0.5) rectangle +(0.4, 1)
            ;
            \draw [yellow, decorate, decoration=snake, double=gray]
                (0.1, \y/3+0.6) -- +(0, 2)
            ;
        }
        \begin{subfigure}[t]{4cm}
            \centering
            \begin{tikzpicture}
                \draw [xshift=0.3cm] (0, 0) -- (3, 0);
                \strand[red]{AGGATTAT}{\prot}
                \strand[blue]{ATTACC}{\prot}
                \strand[teal]{ATTACCATA}{\prot}
                \strand[brown]{GAGAT}{\prot}
            \end{tikzpicture}
            \caption{
                Strands are fixed on a plate.  Ends are protected.
            }
        \end{subfigure}
        \hfill
        \begin{subfigure}[t]{4cm}
            \centering
            \begin{tikzpicture}
                \draw [xshift=0.3cm] (0, 0) -- (3, 0);
                \strand[red]{AGGATTAT}{\prot}
                \strand[blue]{ATTACC}{\light}
                \strand[teal]{ATTACCATA}{\prot}
                \strand[brown]{GAGAT}{\light}
            \end{tikzpicture}
            \caption{
                Shine light on some strands to de-protect their ends.
            }
        \end{subfigure}
        \hfill
        \begin{subfigure}[t]{4cm}
            \centering
            \begin{tikzpicture}
                \begin{scope}
                \strand[red]{AGGATTAT}{\prot}
                \strand[blue]{ATTACC}{}
                \strand[teal]{ATTACCATA}{\prot}
                \strand[brown]{GAGAT}{}
                \end{scope}
                \draw [xshift=0.3cm] (0, 0) -- (3, 0);
                \pgfmathsetseed{20240120}
                \foreach \x in {0, ..., 6}{
                    \foreach \y in{1, ..., 8}{
                        \draw
                            (\x/2+0.3, \y/2)
                            +(rand/4, rand/4) node
                            [fill=white, inner sep=0, rotate=rnd*360]
                            {$\A$}
                        ;
                    }
                }
            \end{tikzpicture}
            \caption{
                The surrounding liquid is filled with
                the next letter, $\A$.
            }
        \end{subfigure}
        \hfill
        \begin{subfigure}[t]{4cm}
            \centering
            \begin{tikzpicture}
                \draw [xshift=0.3cm] (0, 0) -- (3, 0);
                \strand[red]{AGGATTAT}{\prot}
                \strand[blue]{ATTACCA}{}
                \strand[teal]{ATTACCATA}{\prot}
                \strand[brown]{GAGATA}{}
            \end{tikzpicture}
            \caption{
                Floating letters attach to unprotected ends.
                Go to (a).
            }
        \end{subfigure}
        \caption{
            Details of array-based DNA synthesis.
            Each (a)--(d) cycle consumes
            one letter from the master lineup.
            Cf.\ \cite[Fig.~5]{HVS23}.
        }                                             \label{fig:detail}
    \end{figure*}

    DNA molecules are stable and small, making them a good candidate for
    long-term high-density storage.  One challenge DNA storage faces is
    the efficiency of synthesis.  Current technology allows us to
    synthesize multiple DNA strands in parallel to speed up, but with
    one limitation: Different strands cannot be elongated by different
    nucleotides at the same time.  The best we can do is to choose a
    \emph{master lineup}\footnote{ Other works use \emph{synthesis
    sequence}, \emph{reference sequence}, and \emph{reference strand} to
    describe $M$.} $M = M_1 M_2 \dotsm M_t \in \{\A, \C, \G, \T\}^t$
    and, at any given time $s \in \{1, \dotsc, t\}$, ask each of the $n$
    strands if it wants $M_s$ appended at its end.  For illustrations
    of this process, see Fig.~\ref{fig:master}, Fig.~\ref{fig:detail},
    and \cite[Fig.~5]{HVS23} .

    Said limitation raises a mathematical question that, instead of
    trying to synthesize arbitrary $n$-tuples of DNA strands and
    figuring out the optimal $t$, we should fix the time span to be $t$
    and study how many combinations of $n$ strands can be synthesized.

    Such a goal is beyond what
    Lenz et al.\ \cite{LLR20}
    Abu-Sini et al. \cite{ALY23},
    Chrisnata et al.\ \cite{CKL23}, and
    Elishco and Huleihel \cite{ElH23}
    had considered.  Said works are interested in counting the number of
    subsequences of a fixed master lineup; but in our case, master
    lineups are part of the combinatorial objects to be counted.  Our
    questions are also different from
    Makarychev et al.\ \cite{MRR22}
    as they were interested in partitioning desired strands into
    multiple batches---$\{x^1, x^2, \dotsc, x^n\} = B_1 \cup B_2 \cup
    \dotsb \cup B_b$---such that each batch contains ``similar strands''
    that can be synthesized faster together.
    
\subsection{Problem statements}

    The goal of this paper is to answer the following question that is
    based off the assumption that writing and reading DNA are completely
    error-free and the writing time is the only cost we want to optimize
    for.

    \begin{question}                                       \label{que:1}
        How much information can be synthesized into DNA by a machine
        that uses an alphabet $\Sigma$ of size $q$ and can work on $n$
        strands for $t$ units of time.
    \end{question}

    Apart from the fact that half of the fun in math is about seeking
    generalizations, considering arbitrary alphabet size $q$ is
    attractive for several reasons.
    \begin{itemize}
        \item There are nucleobases other than the standard
            $\A$, $\C$, $\G$ and $\T$.
        \item We can use \emph{shortmers} \cite{ALY23}, meaning that we
            can use $\Sigma \coloneqq \{\A\A, \A\C, \dotsc, \T\T\}$ as
            an alphabet of size $16$, or any prefix-free subset $\Sigma
            \subseteq \{\A, \C, \G, \T\}^\star$.
        \item We can use \emph{composite letters} \cite{KYW23}.  A
            composite letter is a distribution on $\{\A, \C, \G, \T\}$,
            for instance $\frac13 \A + \frac23 \T$.
    \end{itemize}

    Question~\ref{que:1} without a doubt is difficult.  We break
    Question~\ref{que:1} into a ladder of questions, starting easy and
    getting harder gradually.

    \begin{question}                                       \label{que:2}
        How many subsequences does the sequence $\A \C \G \T \A \C \G \T
        \A \C \G \T\dotsm$ (repeating $\A$, $\C$, $\G$, and $\T$ until
        there are $t$ letters in total) have?
    \end{question}

    \begin{question}                                       \label{que:3}
        How many pairs $(M, x)$ of sequences are there such that
        $x$ is a subsequence of $M$ and $M$ has length $t$?
    \end{question}

    \begin{question}                                       \label{que:4}
        How many tuples $(M, x^1, \dotsc, x^n)$ of sequences are there
        such that $x^1, \dotsc, x^n$ are subsequences of $M$ and $M$ has
        length $t$?
    \end{question}

    \begin{question}                                       \label{que:5}
        How many ordered tuples $(x^1, \dotsc, x^n)$ of sequences are
        there such that a minimal length supersequence of $x^1, \dotsc,
        x^n$ has length $\leq t$?
    \end{question}

    \begin{question}                                       \label{que:6}
        How many unordered sets $\{x^1, \dotsc, x^n\}$ of sequences are
        there such that a minimal length supersequence of $x^1, \dotsc,
        x^n$ has length $\leq t$?
    \end{question}

    The paper is organized in a way that, for $k \in \{2, 3, 4, 5, 6\}$,
    Section~$k$ answers Question~$k$.

\section{Answering Question~\ref{que:2}}

    We count the subsequences of $\overline{\A\C\G\T}^t = \A \C \G \T \A
    \C \G \T \dotsm$.  Here, the overline repeats the letters below
    until there are $t$ letters.\footnote{ Note that $t$ is not how many
    times it repeats.  This way we can easily express sequences whose
    lengths are not multiples of four.} We denote by $x \preccurlyeq M$
    if $x$ is a subsequence of $M$.

\subsection{The \texorpdfstring{$q = 2$}{q = 2} case}

    Let us begin with counting the number of subsequences of
    $\overline{\A\C}^t$.  Let $\tau$ be a function such that $x
    \preccurlyeq \overline{\A\C}^{\tau(x)}$ but $x \not\preccurlyeq
    \overline{\A\C}^{\tau(x)-1}$ for any binary sequence $x \in \{\A,
    \C\}^\star$.  Think of it as the minimal amount of time required to
    synthesize $x$ given that the master lineup is
    $\overline{\A\C}^\infty$.

    The $\tau$ function has the property that removing the last letter
    of $x$ reduces its $\tau$ by $1$ or $2$.  This constitutes a
    bijection
    \[
        \tau^{-1}(t) \longrightarrow
        \tau^{-1}(t - 1) \cup \tau^{-1}(t - 2)
    \]
    from sequences that need $t$ time to those that need $t - 1$ or $t -
    2$ time.  The bijection gives rise to a recurrence relation
    \[
        |\tau^{-1}(t)| = |\tau^{-1}(t - 1)| + |\tau^{-1}(t - 2)|.
    \]
    Check the initial terms: $|\tau^{-1}(1)| = 1$ and $|\tau^{-1}(2)| =
    2$.  We infer that the number of binary sequences $x \in \{\A,
    \C\}^\star$ such that $\tau(x) = t$ is the $t$th Fibonacci number
    \cite[A000045]{oeis}, which we denote by $F_2(t)$.

    Summing over integers from $0$ to $t$, we infer that the number of
    subsequences of $\overline{\A\C}^t$ is $F_2(0) + \dotsb + F_2(t)$.
    A shorter expression of the partial sum (prefix sum) is $F_2(t + 2)
    - 1$ \cite[A000071]{oeis}.

\subsection{The general \texorpdfstring{$q$}{q} case}

    Suppose that $\Sigma \coloneqq \{a_1, a_2, \dotsc, a_q\}$ is the
    alphabet.  Let $\overline{a_1 a_2 \dotsm a_q}^t$ be the length-$t$
    sequence that cycles through the alphabet like $a_1 a_2 \dotsm a_q
    \, a_1 a_2 \dotsm a_q \, a_1 a_2 \dotsm$.

    The technique of deleting the last letter generalizes to arbitrary
    alphabet size $q$: For any sequence $x \in \Sigma^\star$, let
    $\tau(x)$ be the integer such that $x \preccurlyeq \overline{a_1 a_2
    \dotsm a_q}^{\tau(x)}$ but $x \not\preccurlyeq \overline{a_1 a_2
    \dotsm a_q}^{\tau(x)-1}$.  Removing the last letter of $x$ reduces
    its $\tau$ by at most $q$, creating a bijection 
    \[
        \tau^{-1}(t) \longrightarrow
        \tau^{-1}(t - 1) \cup
        \tau^{-1}(t - 2) \cup \dotsb \cup
        \tau^{-1}(t - q).
    \]
    The bijection translates into a recurrence relation
    \[
        |\tau^{-1}(t)| =
        |\tau^{-1}(t - 1)| +
        |\tau^{-1}(t - 2)| + \dotsb +
        |\tau^{-1}(t - q)|
    \]
    that corresponds to higher-order Fibonacci numbers. 

    It is straightforward to check that $|\tau^{-1}(t)| = 2^{t-1}$ for
    $t \in \{1, \dotsc, q\}$.  As a result, $|\tau^{-1}(t)|$ is the
    $t$th term of the $q$-bonacci numbers, which we denote by $F_q(t)$
    \cite[A048887, A092921]{oeis}.

    \begin{definition} [Higher-order Fibonacci numbers]
        Denote by $F_q(t)$ the $t$th $q$-bonacci number.  It is the sum
        of the last $q$ terms and is $2^{t-1}$ when $t \in \{1, \dotsc,
        q\}$.  $F_2$ is the usual Fibonacci sequence.  $F_3$ is
        sometimes called the Tribonacci sequence \cite[A000073]{oeis}.
    \end{definition}

    The following proposition summarizes what we have so far.

    \begin{proposition}
        The number of the subsequences of $\overline{a_1 \dotsm a_q}^t$
        that are not subsequences of $\overline{a_1 \dotsm a_q}^{t-1}$
        is $F_q(t)$.  The number of the subsequences of $\overline{a_1
        \dotsm a_q}^t$ is the partial sum $F_q(0) + \dotsb + F_q(t)$.
    \end{proposition}

\subsection{Growth of Fibonacci}

    Relating a counting problem to Fibonacci numbers benefits from the
    rich literature.  For instance, using the well-documented generating
    function method, $F_q(t)$ is roughly $\phi_q^t$, where $\phi_q$ is
    the longest\footnote{ Out of the $q + 1$ complex roots of
    \eqref{eq:root}, the \emph{longest} one is the one with the greatest
    modulus.} root of
    \[
        z = 2 - z^{-q}.                                  \label{eq:root}
    \]
    $\phi_q$ happens to be a real number in the interval $[2(1 -
    2^{-q}), 2]$, while the other roots of \eqref{eq:root} all have
    moduli $\leqslant 1$ \cite[Lemma~3.6]{Wol98}.  Plugging
    \eqref{eq:root} into itself, we form a continued fraction expression
    \[
        \phi_q =
        2 - \cfrac{1}{\biggl(\raisebox{1ex}{\scalebox{0.6}{$
        2 - \cfrac{1}{\biggl(\raisebox{1ex}{\scalebox{0.6}{$
        2 - \cfrac{1}{\biggl(\raisebox{1ex}{\scalebox{0.6}{$
        2 - \cfrac{1}{\biggl(\raisebox{1ex}{\scalebox{0.6}{$
        2 - \cfrac{1}{\biggl(\raisebox{1ex}{\scalebox{0.6}{$
        2 - \cfrac{1}{\biggl(\raisebox{1ex}{\scalebox{0.6}{$
        2 - \cfrac{1}{\biggl(\raisebox{1ex}{\scalebox{0.6}{$
        2 - \cfrac{1}{\biggl(\raisebox{1ex}{\scalebox{0.6}{$
        \rule[-1ex]{2em}{1ex}
        $}}\biggr)^q}
        $}}\biggr)^q}
        $}}\biggr)^q}
        $}}\biggr)^q}
        $}}\biggr)^q}
        $}}\biggr)^q}
        $}}\biggr)^q}
        $}}\biggr)^q}                                  \label{for:cfrac}
    \]
    that converges because $2 - z^{-q} - z$ is positive on the interval
    $(1, \phi_q)$ and negative on $(\phi_q, 2)$.

    Earlier works had landed at the same asymptotics.  For instance,
    \cite[Theorem~3]{LLR20} states that the numbers of subsequences of
    $\overline{a_1 \dotsm a_q}^t$ grow as $z_q^{-t}$, where $z_q$ is the
    largest\footnote{ This is a typo; $z_q$ is meant to be the smallest
    root.} root of $z^q + \dotsb + z^1 - 1 = 0$.  Note that $(z - 1)
    (z^q + \dotsb + z^1 - 1) = z^{q+1} - 2z + 1$, which recovers
    \eqref{eq:root}.

\section{Answering Question~\ref{que:3}}

    Let the alphabet size be $|\Sigma| = q$.  We want to express the
    number of pairs $(M, x)$ such that $x \preccurlyeq M \in \Sigma^t$
    as a function in $t$.  We first provide a trivial upper bound.

\subsection{Upper bound}

    \begin{proposition}                                \label{pro:upper}
        The number of pairs $(M, x)$ such that $x \preccurlyeq M \in
        \Sigma^t$ is at most $q^t (F_q(0) + \dotsb + F_q(t))$, where
        $|\Sigma| = q$.
    \end{proposition}

    \begin{IEEEproof}
        $q^t$ is contributed by the choices of $M$.  For any $M$, it
        remains to show that the number of subsequences of $M$ is at
        most $F_q(0) + \dotsb + F_q(t)$.  This is done in the next
        lemma.
    \end{IEEEproof}

    \begin{lemma}                                   \label{lem:maximize}
        Fix alphabet $\Sigma = \{a_1, \dotsc, a_q\}$ and time $t$.
        Master lineups of the form $\overline{a_1 \dotsm a_q}^t$
        maximize the number of subsequences at $F_q(0) + \dotsb +
        F_q(t)$.
    \end{lemma}

    \begin{IEEEproof}
        For any sequence $M \in \Sigma^t$, let $M_{>s}$ be the tail
        sequence $M_{s+1} M_{s+2} \dotsm M_t$ obtained by ignoring the
        first $s$ letters.  We observe that, for a subsequence $x
        \preccurlyeq M$ that begins with $x_1 = a_1 \in \Sigma$, the
        rest of $x$, i.e., $x_{>1}$, must be a subsequence of
        $M_{>f_1}$, where $f_1$ is the index of the first occurrence of
        $a_1$ in $M$, i.e., $f_1 \coloneqq \min(\{s \,|\, M_s{=}a_1\}
        \cup \{\infty\})$.
        
        Denote by $[M]$ the set of subsequences of $M$.  Deleting the
        first letter constitutes a bijection
        \[
            [M] \longrightarrow
            [M_{>f_1}] \sqcup
            [M_{>f_2}] \sqcup \dotsb \sqcup
            [M_{>f_q}]
        \]
        where $f_u$ is the index of the first $a_u$ in $M$, $u \in \{1,
        \dotsc, q\}$, and $\sqcup$ denotes disjoint union\footnote{
        Unlike union, disjoint union does not merge common elements,
        e.g., $|\{1, 2\} \sqcup \{2, 3\}| = 4$, not $3$.}.  This
        corresponds to a recurrence relation
        \[
            |[M]| =
            |[M_{>f_1}]| +
            |[M_{>f_2}]| + \dotsb +
            |[M_{>f_q}]|
        \]
        which implies suboptimality
        \[
            |[M]| \leqslant
            |[M_{>1}]| +
            |[M_{>2}]| + \dotsb +
            |[M_{>q}]|
            \label{ine:subop}
        \]
        unless $f_1, \dotsc, f_q$ is a permutation of $1, \dotsc, q$.
    \end{IEEEproof}

    The fact that aperiodic master lineups maximize the number of
    subsequences has been observed by previous works.  For instance,
    Calabi and Hartnett stated in \cite[Lemma~5.7]{CaH69} that they
    maximize the number of subsequences for any given length.  See
    \cite[Theorem~1]{Hir99} and \cite[Theorem~2.4]{HiR00}
    \cite[Corollary~2]{LiL15} for proofs.  Summing
    \cite[Lemma~5.7]{CaH69} over all possible lengths, we recover
    Lemma~\ref{lem:maximize}.

    Moreover, \cite[Lemma~5.7]{CaH69} expresses the number of
    subsequences of $\overline{\A\C}^t$ of length $t - r$, where $r$ can
    be understood as the deletion radius, as
    \[ \sum_{i=0}^r \binom{n-r}{i}. \]
    This can be seen as summing entries of the Pascal triangle. As
    illustrated in Figure~\ref{fig:pascal}, this is compatible with the
    expression of Fibonacci numbers on the Pascal triangle.  We thank an
    anonymous reviewer for pointing out the connection.

    Next, we want to bound the number of pairs $(M, x)$ from below.  The
    lower bound will deviate significantly from the upper bound because
    a nontrivial amount of subsequences are lost in inequality
    \eqref{ine:subop} when we upper bound $|[M_{>f_1}]| + \dotsb +
    |[M_{>q}]|$ by $|[M_{>1}]| + \dotsb + |[M_{>q}]|$.  One can expect
    the final answer to be some nontrivial exponential-growth function
    between $\phi_q^t$ and $(q \phi_q)^t$.

    \begin{figure}
        \begin{tikzpicture}
            \foreach \t in {0, ..., 7} {
                \draw [yellow!80!black, line width=2ex]
                    (-\t/4 - 1/4, -\t/2) --
                    ({min(\t/4, ((7 - \t)/2 - \t/4)) + 1/4}, -\t/2)
                ;
            }
            \foreach \t in {0, ..., 7} {
                \foreach \r in {0, ..., \t} {
                    \draw (\r/2 - \t/4, -\t/2) node {$\binom\t\r$};
                }
            }
        \end{tikzpicture}
        \hfill
        \begin{tikzpicture}
            \foreach \t in {0, ..., 7} {
                \draw [yellow!80!black, line width=1ex]
                    (-\t/4 - 3/8, -\t/2 - 1/4) --
                    +(\t*3/8 + 6/8, \t/4 + 2/4)
                ;
            }
            \foreach \t in {0, ..., 7} {
                \foreach \r in {0, ..., \t} {
                    \draw (\r/2 - \t/4, -\t/2) node {$\binom\t\r$};
                }
            }
        \end{tikzpicture}
        \caption{
            Left: \cite[Lemma~5.7]{CaH69} counts the subsequences of
            $\overline{\A\C}^t$ by length.  Right: we count by $\tau$
            (i.e., synthesis time).
        }                                             \label{fig:pascal}
    \end{figure}

\subsection{Lower bound for \texorpdfstring{$q = 3$}{q = 3}}

    Consider $q = 3$ as an example.  We want to focus on master lineups
    $M \in \{\A, \C, \G\}^t$ that do not contain $\A\A$, $\C\C$, or
    $\G\G$.  There are $3 \cdot 2^{t-1}$ such sequences.  Now we lower
    bound the number of subsequences.

    \begin{lemma}
        If an $M \in \{\A, \C, \G\}^t$ does not contain $\A\A$, $\C\C$,
        or $\G\G$, the number of its subsequences is $\geqslant F_2(0) +
        \dotsb + F_2(t) = F_2(t + 2) - 1$.
    \end{lemma}

    \begin{IEEEproof}
        Let $M_1 M_2 \dotsm M_t$ be the content of $M$.  Let $\tau(x)$
        be the least integer such that $x$ is a subsequence of $M_1 M_2
        \dotsm M_{\tau(x)}$.  By deleting the last letter of a
        subsequence, we construct a bijection
        \[
            \tau^{-1}(t) \longrightarrow
            \bigcup_{u=1}^{\text{until }M_{t-u}=M_t}
            \tau^{-1}(t - u),                          \label{bij:until}
        \]
        where the union stops after the first offset $u$ such that
        $M_{t-u}$ matches the letter being deleted, $M_t$.

        We have assumed that $M$ is free of $\A\A$, $\C\C$, or $\G\G$ so
        it is guaranteed that $M_{t-1} \neq M_t$.  This means that the
        $u$ in \eqref{bij:until} must at least run over $1$ and $2$,
        implying the recurrence relation
        \[
            |\tau^{-1}(t)| \geqslant
            |\tau^{-1}(t - 1)| + |\tau^{-1}(t - 2)|.   \label{ine:until}
        \]
        With initial values $|\tau^{-1}(1)| = 1$ and $|\tau^{-1}(2)| =
        2$, we conclude that $|\tau^{-1}(t)| \geqslant F_2(t)$.  This
        proves the lemma after taking the partial sum.
    \end{IEEEproof}

    We now combine the two facts---that there are $3 \cdot 2^{t-1}$
    master lineups that avoid $\A\A$, $\C\C$, and $\G\G$, and each of
    them has at least $F_2(t + 2) - 1$ subsequences.

    \begin{proposition}                                 \label{pro:3.24}
        The number of pairs $(M, x)$ such that $x \preccurlyeq M \in
        \{\A, \C, \G\}^t$ is at least $3 \cdot 2^{t-1} \cdot (F_2(t + 2)
        - 1)$.
    \end{proposition}

    The main term of Proposition~\ref{pro:3.24} is $2^t \phi_2^t \approx
    3.24^t$.  This is greater than the number of subsequences of
    $\overline{\A\C\G}^t$, which is about $\phi_3^t \approx 1.84^t$.  On
    the other hand, we still are a distance away from the upper bound
    $(3\phi_3)^t \approx 5.51^t$.

\subsection{Lower bound for general \texorpdfstring{$q$}{q}}

    For general $q$, one can follow the same recipe to prove the
    following generalization of Proposition~\ref{pro:3.24}.

    \begin{proposition}                                 \label{pro:u<=2}
        Fix $|\Sigma| = q$ and a time $t$.  The number of pairs $(M, x)$
        such that $x \preccurlyeq M \in \Sigma^t$ is greater than or
        equal to $q (q - 1)^{t-1} \cdot (F_2(t + 2) - 1)$.
    \end{proposition}

    \begin{IEEEproof}
        Use master lineups that do not contain $a_1a_1$, $a_2a_2$, ...,
        or $a_qa_q$.  The number of such master lineups is no less than
        $q (q - 1)^{t-1}$.  The number of subsequences this type of
        master lineups have is no less than $F_2(t + 2) - 1$ due to a
        bijection that looks like \eqref{bij:until} and a recurrence
        relation that look like \eqref{ine:until}.
    \end{IEEEproof}

    Proposition~\ref{pro:u<=2} can be further generalized by considering
    how many preimages we want to unionize in \eqref{bij:until}: In the
    last subsection, we assume $M_s \neq M_{s-1}$.  We can add one more
    restriction $M_s \neq M_{s-2}$, which allows $u$ run through $1$,
    $2$, and $3$.  This adds an extra term $|\tau^{-1}(t - 3)|$ into
    \eqref{ine:until}, making $|\tau^{-1}(t)|$ greater than or equal to
    the Tribonacci number.

    Adding more terms into \eqref{ine:until} comes with a price---there
    are fewer master lineups that also satisfy $M_s \neq M_{s-2}$ for
    all $s \in \{3, \dotsc, t\}$.  More quantitatively, there are at
    least $(q - 2)^t$ master lineups such that $M_s \notin \{M_{s-1},
    M_{s-2}\}$ for all $s$.  This yields the following result.

    \begin{proposition}                                 \label{pro:u<=3}
        Fix $|\Sigma| = q$ and a time $t$.  The number of pairs $(M, x)$
        such that $x \preccurlyeq M \in \Sigma^t$ is greater than or
        equal to $(q - 2)^t (F_3(0) + \dotsb + F_3(t))$.
    \end{proposition}

    The main term of Proposition~\ref{pro:u<=3} is $((q - 2) \phi_3)^t$,
    where $\phi_3 \approx 1.84$.  Up to some constant scalars, the same
    technique provides lower bounds $((q - 3) \phi_4)^t$, $((q - 4)
    \phi_5)^t$, and so on.

    \begin{proposition}
        Let $p$ be an integer, $2 \leqslant p \leqslant q$.  Let
        $|\Sigma| = q$.  The number of tuples $(M, x)$ such that $x
        \preccurlyeq M \in \Sigma^t$ is at least $(q + 1 - p)^t (F_p(0)
        + \dotsb + F_p(t))$.
    \end{proposition}

    Because $\phi_p \approx 2 - 2^{-p}$ (cf.\ \eqref{for:cfrac}), the
    optimal lower bound that can be obtained this way is somewhere
    around $p \approx \log q$, yielding $((q - \log q) \phi_{\log q})^t
    = (2q - O(\log q))^t$.  Compare this to the upper bound $(q\phi_q)^t
    = (2q - O(2^{-q}))^t$ and the trivial lower bound $\phi_q^t$
    obtained by fixing a master lineup.

\section{Answering Question~\ref{que:4}}

    Fix alphabet size $|\Sigma| = q$, time $t$, and number of strands
    $n$ a DNA machine can handle.  We now want to count the tuples $(M,
    x^1, \dotsc, x^n)$ such that $x^1, \dotsc, x^n \preccurlyeq M \in
    \Sigma^t$.

\subsection{The bounds}

    \begin{proposition}                                 \label{pro:free}
        Let $p$ be an integer, $2 \leqslant p \leqslant q$.  Let
        $|\Sigma| = q$.  The number of tuples $(M, x^1, \dotsc, x^n)$
        such that $x^1, \dotsc, x^n \preccurlyeq M \in \Sigma^t$ is at
        most $q^t (F_q(0) + \dotsb + F_q(t))^n$ but at least $(q + 1 -
        p)^t (F_p(0) + \dotsb + F_p(t))^n$.
    \end{proposition}

    \begin{IEEEproof}
        For the upper bound, the same idea of
        Proposition~\ref{pro:upper} applies.  We skip the details.

        For lower bounds, the idea is to use the free parameter $p$ to
        control how many preimages we want to unionize in
        \eqref{bij:until} and how many summands are in
        \eqref{ine:until}.  We consider master lineups $M$ wherein any
        $p$ consecutive letters $M_s \dotsm M_{s+p-1}$ are all distinct.
        The number of this type of master lineups is at least $(q + 1 -
        p)^t$.

        Now that $M_{t-p+1} \dotsm M_t$ are all distinct, the $u$ in
        \eqref{bij:until} will run from $1$ to at least $p$.  This
        implies
        \[
            |\tau^{-1}(t)| \geqslant
            |\tau^{-1}(t - 1)| + \dotsb + |\tau^{-1}(t - p)|.
        \]
        Thus $\tau^{-1}(t)$ is at least $F_p(t)$.  Take the partial sum
        to lower bound the number of subsequences of $M$ by $F_p(0) +
        \dotsb + F_p(t)$.  Because this proposition concerns $n$-tuples
        of subsequences, the lower bound becomes $(F_p(0) + \dotsb +
        F_p(t))^n$.  This completes the proof of the lower bound.
    \end{IEEEproof}

\subsection{The best \texorpdfstring{$p$}{p}}

    The lower bound in Proposition~\ref{pro:free} contains a free
    parameter $p$.  we study the optimal $p$ in terms of $q$, $t$, and
    $n$.

    First we use $\phi_p^t$ to approximate the Fibonacci number.  The
    lower bound becomes $(q + 1 - p)^t \phi_p^{tn}$.  Now that both
    factors have $t$ in exponents, $t$ does not affect the optimal value
    of $p$.  It remains to maximize $(q + 1 - p) \phi_p^n$ $\approx (q +
    1 - p) (2 - 2^{-p})^n$.  From here we see that the best $p$ should
    be around $\log(qn)$ and defaults to $q$ when $n$ is about or beyond
    the order of $2^q$.

\section{Answering Question~\ref{que:5}}

    Question~\ref{que:5} goes one step ahead of Question~\ref{que:4} in
    difficulty.  As we no longer ``remember'' the master lineup, the
    information encoded in $M$ can only be recovered if somehow the
    subsequences uniquely determine $M$.

\subsection{The intuition}

    Our idea is based off the greedy algorithm for finding a common
    supersequence of $x^1, \dotsc, x^n$.
    \begin{itemize}
        \item Start from an empty sequence $M \coloneqq$.
        \item Observe the first letters of $x^1, \dotsc, x^n$.
        \item If $a \in \Sigma$ is the most common first letter, append
            $a$ to $M$ and delete the starting $a$ in those $x$'s that
            do start with $a$.
        \item \tikz[overlay]\draw[->, red!90!black, line width=1pt]
            (-1em, 0) to [out=165, in=195] (-1em, 4em);%
            Repeat until $x^1, \dotsc, x^n$ deplete.
    \end{itemize}
    We turn the greedy algorithm around and make it a design principle:
    We only count the tuples $(x^1, \dotsc, x^n)$ of subsequences of $M$
    that recover $M$ under the greedy algorithm.

    In order to guarantee that $M_1$ will be the most common first
    letter among $x^1, \dotsc, x^n$, we will enforce that more than half
    of the subsequences will begin with $M_1$.  And then, once the
    starting $M_1$ is removed from the subsequences, we will enforce
    that more than half of the subsequences will begin with $M_2$.  The
    same criteria is applied to every $M_s$, $s \in \{1, \dotsc, t\}$.

    We use a binary matrix $Y \in \{0, 1\}^{t\times n}$ to encode the
    two criteria we want the subsequences to have.
    \begin{itemize}
        \item [(a)] The $s$th row of $Y$ is the indicator of which
            strands want $M_s$.  We want that half of the row is one.
            (For simplicity, assume that $n$ is odd.)
        \item [(b)] The $j$th column of $Y$ is the indicator of which
            letters $x^j$ inherits from $M$.  We want that there is no
            consecutive zeros of length $p$ (to avoid over-counting
            $x^j$).
    \end{itemize}
    We want to compute the number of matrices that satisfy both (a) and
    (b).  We notice that both criteria are monotonic.  For example, if a
    row of $Y$ contains more ones than there are zeros, then (a) remains
    true after a $0$ turns into a $1$.  This allows application of the
    FKG inequality, which states that increasing functions are
    positively correlated.  (FKG assumes that the underlying probability
    measure is log-supermodular; in our case, product measure is always
    log-supermodular.)

    \begin{lemma}[FKG inequality]
        If $f$ and $g$ are increasing functions and the entries of $Y$
        are iid uniform on $\{0, 1\}$, then $\EE[f(Y)g(Y)] \geqslant
        \EE[f(Y)] \EE[g(Y)]$.  \cite[Proposition~1']{FKG71}
    \end{lemma}

    With FKG we can lower bound the number of ``master-less'' tuples,
    tuples that consist of $M$'s subsequences but do not remember $M$
    explicitly.

\subsection{The formal proof}

    \begin{theorem}                                     \label{thm:no-M}
        Let $n$ be odd.  Let $p$ be an integer, $2 \leqslant p \leqslant
        q$.  The number of tuples $(x^1, \dotsc, x^n)$ such that $x^1,
        \dotsc, x^n \preccurlyeq M$ for some $M \in \Sigma^t$, where
        $|\Sigma| = q$, is bounded from below by $(q + 1 - p)^t F_p(t +
        1)^n / 2^t$.
    \end{theorem}

    \begin{IEEEproof}
        We continue using master lineups $M$ that avoid repeated letter
        in any $p$ consecutive letters.  The number of such master
        lineups is $\geqslant (q + 1 - p)^t$.

        For each such $M$, we want to count the number of tuples $(x^1,
        \dotsc, x^n)$.  This is equivalent to counting the number of
        binary matrices $Y$ that satisfy both (a) and (b).

        Let $f\colon \{0, 1\}^{t\times n} \to \{0, 1\}$ be the indicator
        of whether or not all columns avoid $p$ consecutive zeros.  More
        precisely,
        \[
            f(Y) \coloneqq \prod_{k=1}^n \prod_{s=0}^{t-p}
            \min(1, Y_{s+1,k} + \dotsb + Y_{s+p,k})
        \]
        Let the entries of $Y$ be iid uniform on $\{0, 1\}$.  We know
        \[
            \EE \biggl[
                \prod_{s=0}^{t-p}
                \min(1, Y_{s+1,k} + \dotsb + Y_{s+p,k})
            \biggr]
            \geq \frac{F_p(t + 1)}{2^t}.
        \]
        This is because the number of binary sequences that avoid $p$
        consecutive zeros is the $p$-bonacci number and $2^t$ is the
        total number of binary sequences.  From there we infer
        \[ \EE[f(Y)] \geqslant \frac {F_p(t + 1)^n} {2^{tn}} \]
        because each column of $Y$ are drawn independently.

        Let $g\colon \{0, 1\}^{t\times n} \to \{0, 1\}$ be the indicator
        of whether or not all rows have more ones than zeros.  More
        precisely,
        \[
            g(Y) \coloneqq \prod_{s=1}^t
            \frac{1 + \operatorname{sgn}
            (Y_{s,1} + \dotsb + Y_{s,n} - n/2)}{2}.
        \]
        We assume $n$ odd and don't worry about ties.  Each row has more
        ones than zeros with probability $1/2$.  As a consequence,
        \[ \EE[g(Y)] \geqslant \frac{1}{2^t}. \]
        Apply FKG,
        \[
            \EE[f(Y)g(Y)] \geqslant
            \EE[f(Y)] \EE[g(Y)] \geqslant
            \frac {F_p(t+1)^n} {2^{tn+t}}.
        \]
        Subsequently, the number of valid $Y$'s is greater than or equal
        to $F_p(t + 1)^n / 2^t$.  This quantity multiplied by the number
        of master lineups completes the proof of the theorem.
    \end{IEEEproof}

    Theorem~\ref{thm:no-M} implies that the number of master-less tuples
    is about $\phi_p^{(t+1)n} / 2^t$, if not more.

\section{Answering Question~\ref{que:6}}

    Although synthesizing multiple DNA strands in parallel saves time,
    current technology does not allow us to separate $x^1, \dotsc, x^n$
    after the process completes.  The strands will float freely in a
    liquid container and a DNA reader will almost always read the
    strands out of order.

    Because this paper does not plan to address how to encode and decode
    the order on $x^1, \dotsc, x^n$, we simply divide the count in
    Theorem~\ref{thm:no-M} by $n!$ in the following corollary.

    \begin{corollary}
        Let $n$ be odd.  Assume $2 \leqslant p \leqslant q$.  The number
        of unordered sets $\{x^1, \dotsc, x^n\}$ such that $x^1, \dotsc,
        x^n \preccurlyeq M$ for some $M \in \Sigma^t$, where $|\Sigma| =
        q$, is bounded from below by $(q + 1 - p)^t F_p(t + 1)^n / (2^t
        n!)$.
    \end{corollary}

    We close our paper with an upper bound.

    \begin{proposition}
        Let $t$, $n$, and $q$ be positive integers.  Let $\Sigma$ be of
        size $q$.  The number of unordered sets $\{x^1, \dotsc, x^n\}$
        such that $x^1, \dotsc, x^n \preccurlyeq M$ for some $M \in
        \Sigma^t$ is bounded from above by $q^t \binom{F_q(0) + \dotsb +
        F_q(t)}{\leqslant n}$.
    \end{proposition}

\section{Conclusion}

    In this paper, we study how many combinations a DNA synthesis
    machine can create, given that the machine uses an alphabet of size
    $q$, works on $n$ strands in parallel, and lasts $t$ time.  We
    breakdown this big challenge into a series of smaller puzzles, make
    small breakthroughs at each step, and synthesize the final answer at
    the end of the paper.

    Here are some highlights of our techniques.  In answering
    Question~\ref{que:2}, we relate the subsequence problem to Fibonacci
    numbers.  In answering Question~\ref{que:3}, we propose a scheme
    that encodes information into the choices of $M$, breaking the
    trivial lower bound that assumes a fixed $M$.  In answering
    Question~\ref{que:4}, we ``remember'' $M$ as part of the tuples,
    paving the way for the next question.  In answering
    Question~\ref{que:5}, we ``forget'' $M$ but use greedy algorithm to
    recover $M$ upon seeing $x^1, \dotsc, x^n$.  Finally, we put
    together everything to answer Question~\ref{que:6}, which is the
    quantitative version of Question~\ref{que:1}.

\IEEEtriggeratref{7}
\bibliographystyle{IEEEtran}
\bibliography{FiboNotSeen-26}

\begin{thebibliography}{10}
\providecommand{\url}[1]{#1}
\csname url@samestyle\endcsname
\providecommand{\newblock}{\relax}
\providecommand{\bibinfo}[2]{#2}
\providecommand{\BIBentrySTDinterwordspacing}{\spaceskip=0pt\relax}
\providecommand{\BIBentryALTinterwordstretchfactor}{4}
\providecommand{\BIBentryALTinterwordspacing}{\spaceskip=\fontdimen2\font plus
\BIBentryALTinterwordstretchfactor\fontdimen3\font minus \fontdimen4\font\relax}
\providecommand{\BIBforeignlanguage}[2]{{%
\expandafter\ifx\csname l@#1\endcsname\relax
\typeout{** WARNING: IEEEtran.bst: No hyphenation pattern has been}%
\typeout{** loaded for the language `#1'. Using the pattern for}%
\typeout{** the default language instead.}%
\else
\language=\csname l@#1\endcsname
\fi
#2}}
\providecommand{\BIBdecl}{\relax}
\BIBdecl

\bibitem{HVS23}
\BIBentryALTinterwordspacing
A.~Hoose, R.~Vellacott, M.~Storch, P.~S. Freemont, and M.~G. Ryadnov, ``{{DNA}} synthesis technologies to close the gene writing gap,'' \emph{Nature Reviews Chemistry}, vol.~7, no.~3, pp. 144--161, Jan. 2023. [Online]. Available: \url{https://www.nature.com/articles/s41570-022-00456-9}
\BIBentrySTDinterwordspacing

\bibitem{LLR20}
\BIBentryALTinterwordspacing
A.~Lenz, Y.~Liu, C.~Rashtchian, P.~H. Siegel, A.~{Wachter-Zeh}, and E.~Yaakobi, ``Coding for {{Efficient DNA Synthesis}},'' in \emph{2020 {{IEEE International Symposium}} on {{Information Theory}} ({{ISIT}})}.\hskip 1em plus 0.5em minus 0.4em\relax {Los Angeles, CA, USA}: {IEEE}, Jun. 2020, pp. 2885--2890. [Online]. Available: \url{https://ieeexplore.ieee.org/document/9174272/}
\BIBentrySTDinterwordspacing

\bibitem{ALY23}
\BIBentryALTinterwordspacing
M.~{Abu-Sini}, A.~Lenz, and E.~Yaakobi, ``{{DNA Synthesis Using Shortmers}},'' in \emph{2023 {{IEEE International Symposium}} on {{Information Theory}} ({{ISIT}})}.\hskip 1em plus 0.5em minus 0.4em\relax {Taipei, Taiwan}: {IEEE}, Jun. 2023, pp. 585--590. [Online]. Available: \url{https://ieeexplore.ieee.org/document/10206609/}
\BIBentrySTDinterwordspacing

\bibitem{CKL23}
\BIBentryALTinterwordspacing
J.~Chrisnata, H.~M. Kiah, and V.~Long Phuoc~Pham, ``Deletion {{Correcting Codes}} for {{Efficient DNA Synthesis}},'' in \emph{2023 {{IEEE International Symposium}} on {{Information Theory}} ({{ISIT}})}.\hskip 1em plus 0.5em minus 0.4em\relax {Taipei, Taiwan}: {IEEE}, Jun. 2023, pp. 352--357. [Online]. Available: \url{https://ieeexplore.ieee.org/document/10206892/}
\BIBentrySTDinterwordspacing

\bibitem{ElH23}
\BIBentryALTinterwordspacing
O.~Elishco and W.~Huleihel, ``Optimal {{Reference}} for {{DNA Synthesis}},'' \emph{IEEE Transactions on Information Theory}, vol.~69, no.~11, pp. 6941--6955, Nov. 2023. [Online]. Available: \url{https://ieeexplore.ieee.org/document/10155240/}
\BIBentrySTDinterwordspacing

\bibitem{MRR22}
\BIBentryALTinterwordspacing
K.~Makarychev, M.~Z. Racz, C.~Rashtchian, and S.~Yekhanin, ``Batch {{Optimization}} for {{DNA Synthesis}},'' \emph{IEEE Transactions on Information Theory}, vol.~68, no.~11, pp. 7454--7470, Nov. 2022. [Online]. Available: \url{https://ieeexplore.ieee.org/document/9801834/}
\BIBentrySTDinterwordspacing

\bibitem{KYW23}
\BIBentryALTinterwordspacing
A.~Kobovich, E.~Yaakobi, and N.~Weinberger, ``M-{{DAB}}: {{An Input-Distribution Optimization Algorithm}} for {{Composite DNA Storage}} by the {{Multinomial Channel}},'' 2023. [Online]. Available: \url{http://arxiv.org/abs/2309.17193}
\BIBentrySTDinterwordspacing

\bibitem{oeis}
{OEIS Foundation Inc.}, ``The {O}n-{L}ine {E}ncyclopedia of {I}nteger {S}equences,'' 2024, published electronically at \url{http://oeis.org}.

\bibitem{Wol98}
D.~Wolfram, ``Solving generalized {{Fibonacci}} recurrences,'' \emph{Fibonacci Quarterly}, vol.~36, pp. 129--145, May 1998.

\bibitem{CaH69}
\BIBentryALTinterwordspacing
L.~Calabi and W.~Hartnett, ``Some general results of coding theory with applications to the study of codes for the correction of synchronization errors,'' \emph{Information and Control}, vol.~15, no.~3, pp. 235--249, Sep. 1969. [Online]. Available: \url{https://linkinghub.elsevier.com/retrieve/pii/S0019995869904422}
\BIBentrySTDinterwordspacing

\bibitem{Hir99}
\BIBentryALTinterwordspacing
D.~S. Hirschberg, ``Bounds on the {{Number}} of {{String Subsequences}},'' in \emph{Combinatorial {{Pattern Matching}}}, M.~Crochemore and M.~Paterson, Eds.\hskip 1em plus 0.5em minus 0.4em\relax Berlin, Heidelberg: Springer Berlin Heidelberg, 1999, vol. 1645, pp. 115--122. [Online]. Available: \url{http://link.springer.com/10.1007/3-540-48452-3_9}
\BIBentrySTDinterwordspacing

\bibitem{HiR00}
D.~S. Hirschberg and M.~Regnier, ``Tight {{Bounds}} on the {{Number}} of {{String Subsequences}},'' 2000.

\bibitem{LiL15}
\BIBentryALTinterwordspacing
Y.~Liron and M.~Langberg, ``A {{Characterization}} of the {{Number}} of {{Subsequences Obtained}} via the {{Deletion Channel}},'' \emph{IEEE Transactions on Information Theory}, vol.~61, no.~5, pp. 2300--2312, May 2015. [Online]. Available: \url{http://ieeexplore.ieee.org/document/7061929/}
\BIBentrySTDinterwordspacing

\bibitem{FKG71}
\BIBentryALTinterwordspacing
C.~M. Fortuin, P.~W. Kasteleyn, and J.~Ginibre, ``Correlation inequalities on some partially ordered sets,'' \emph{Communications in Mathematical Physics}, vol.~22, no.~2, pp. 89--103, Jun. 1971. [Online]. Available: \url{http://link.springer.com/10.1007/BF01651330}
\BIBentrySTDinterwordspacing

\end{thebibliography}

\end{document}